\documentclass[12pt]{article}

\usepackage[utf8]{inputenc} % allow utf-8 input
\usepackage[T1]{fontenc}    % use 8-bit T1 fonts
\usepackage{hyperref}       % hyperlinks

\usepackage{longtable}
\usepackage{booktabs}
\usepackage{multirow}

\usepackage{caption}
\usepackage{subcaption}
\usepackage{graphicx}

\usepackage{amsfonts,amsmath,amssymb}
\usepackage{nicefrac}

\renewcommand{\Pr}{\operatorname{Pr}}
\newcommand{\E}{\ensuremath{\operatorname{E}}}
\newcommand{\V}{\operatorname{V}}
\newcommand{\st}{\operatorname{subject\ to:\ }}
\newcommand{\argmin}[1]{\operatornamewithlimits{argmin:\ }_{#1}}

\newcommand{\CP}{\ensuremath{\operatorname{CP}}}
\newcommand{\ACP}{\ensuremath{\operatorname{ACP}}}
\newcommand{\OCP}{\ensuremath{\operatorname{OCP}}}
\newcommand{\PP}{\ensuremath{\operatorname{PP}}}
\newcommand{\EP}{\ensuremath{\operatorname{EP}}}
\renewcommand{\Pr}{\ensuremath{\operatorname{Pr}}}
\newcommand{\cond}{\ensuremath{\,|\,}}
\usepackage{color}
\usepackage{hyperref}
\usepackage[numbers,super,sort&compress]{NJDnatbib}
\usepackage{scalerel}

\usepackage{comment}

\title{Conditional Power and Friends: The Why and How of (Un)planned, Unblinded Sample Size Recalculations in Confirmatory Trials}

\author{
    Kevin Kunzmann \thanks{
        MRC Biostatistics Unit, University of Cambridge,
        Cambridge Institute of Public Health,
        Forvie Site, Robinson Way,
        Cambridge Biomedical Campus,
        Cambridge CB2 0SR,
        United Kingdom,
        \texttt{kevin.kunzmann@mrc-bsu.cam.ac.uk}
    }\\
    MRC Biostatistics Unit \\
    University of Cambridge \\
    Michael J. Grayling \\
    Population Health Sciences Institute \\
    Newcastle University \\
    Kim May Lee \\
    Pragmatic Clinical Trials Unit\\
    Queen Mary University of London\\
    David S.\ Robertson \\
    MRC Biostatisics Unit\\
    University of Cambridge\\
    Kaspar Rufibach \\
    Methods, Collaboration, and Outreach Group (MCO) \\
    Department of Biostatistics\\
    F. Hoffmann-La Roche, Basel\\
    James M. S. Wason \\
    Population Health Sciences Institute \\
    Newcastle University\\
    and \\
     MRC Biostatistics Unit \\
    University of Cambridge
}
\usepackage{setspace}
\begin{document}

\maketitle

\captionsetup{width=\textwidth}

%\newpage
%\section*{Funding}
%\noindent
%David S. Robertson was funded by the Biometrika Trust and the Medical Research %Council under Grant MC\_UU\_00002/6.

\newpage

\begin{abstract}
Adapting the final sample size of a trial to the evidence accruing
during the trial is a natural way to address planning uncertainty.
Designs with adaptive sample size need to
account for their optional stopping to guarantee
strict type-I error-rate control.
A variety of different methods to maintain type-I error-rate control
after unplanned changes of the initial sample size have been proposed
in the literature.
This makes interim analyses for the purpose of sample size
recalculation feasible in a regulatory context.
Since the sample size is usually determined via an
argument based on the power of the trial,
an interim analysis raises the question of
how the final sample size should be determined conditional on the accrued information.
Conditional power is a concept often put forward in this context.
Since it depends on the unknown effect size,
we take a strict estimation perspective and compare
assumed conditional power, observed conditional power,
and predictive power with respect to their properties as estimators
of the unknown conditional power.
We then demonstrate that pre-planning an interim analysis using
methodology for unplanned interim analyses is ineffective and
naturally leads to the concept of optimal two-stage designs.
We conclude that unplanned design adaptations should only be conducted
as reaction to trial-external new evidence, operational needs to violate the originally chosen design,
or \textit{post~hoc} changes in the objective criterion.
Finally, we show that commonly discussed sample size recalculation rules can lead to paradoxical outcomes and propose two alternative ways of reacting to
newly emerging trial-external evidence.
\end{abstract}


\newpage

\section{Introduction}
\label{sec:introduction}

The planning phase of confirmatory clinical trials is typically
characterized by substantial uncertainty about the magnitude of the
parameters underlying the hypothesis of interest.
Often the alternative hypothesis is the superiority of a new intervention arm
over a treatment-as-usual control arm and is formalized via
the average treatment effect, e.g., the difference of means between the
two randomly allocated treatment arms.
Clinical data collection is expensive and time consuming leading to a
strong economic incentive to reach the study goals with as little data as possible.
The conventional statistical criteria to determine the sample size of a trial
are a one-sided type-I error rate of
2.5\% and a power of 80\% or 90\%.
Since power is a function of the unknown effect size the initial design
must be specified under substantial uncertainty about the magnitude of the effect which the trial hopes to detect eventually.
Mainly, two ways of addressing this challenge have been put forward in the literature.

Firstly, the so-called `hybrid' approach to sample size
derivation takes a Bayesian
view on determining the initial sample size of a clinical trial~\cite{spiegelhalter1994}.
The Bayesian component requires the specification of an informative
prior on the parameters of interest.
This then allows reasoning about the `expected power' of a trial as a
function of its sample size and, consequently, determination of the sample size
such that the expected power exceeds a target threshold.
This concept is a straight-forward extension of the usual practice of
computing the power under a fixed point alternative, which can easily be recovered as a special case when considering a point prior.
The advantage lies in the fact that the \textit{a~priori} information about the magnitude of the effect size is faithfully
reflected in the sample size derivation.
Since the actual analysis is still frequentist, type-I error rate control is not compromised.

Secondly, authors have proposed to apply the concept of adaptive design changes
to recalculate the initial sample size of a design based on data observed
within the trial itself~\cite{bauer2016}.
The rationale behind this approach is to use the accruing evidence about the unknown parameters driving the sample size derivation to
`correct' the sample size mid-trial.
During the interim analysis, the accrued data can either be unblinded
or not.
Methods for the latter `blinded interim analysis' are particularly useful if relevant nuisance parameters such as the variance are also unknown \cite{birkett1994internal,bauer2016}.
We focus on the unblinded case which allows for a more precise
interim assessment of the effect size than methods that retain
the blind.
The maximal type-I error rate constraint is then usually protected by
applying the conditional error principle~\cite{muller2004,brannath2012}
or an equivalent formulation via $\alpha$-spending or
combination functions~\cite{bauer2016}.
Often the sample size of the current
trial is adjusted such that the conditional power given the data observed up to the
interim analysis again exceeds the initial threshold for unconditional
power~\cite{proschan1995}.
Any recalculation based on conditional power arguments must
address the problem that conditional power,
just as unconditional power,
depends on the unknown underlying parameters.
It must thus be
estimated to inform a sample size recalculation.
Yet, precise estimation of the relevant parameters before the conclusion of a trial is hard since only a fraction of the final sample size is available.
Three approaches to estimating conditional power have been discussed in the literature.

Firstly,
`conditional power' is the probability to reject the null hypothesis given the interim data as a function of the unknown parameters.
In a slight abuse of terminology, the same term is often used to refer to
the assumed conditional power that is obtained by plugging in the
point alternative used for the initial sample size derivation.
Evidently, this assumed conditional makes no use of the accrued trial
data since the effect size is kept fixed.
Secondly,
authors have proposed to use the `observed conditional power' instead, which replaces the unknown parameters with their maximum-likelihood estimates.
The latter `plug-in' approach is often criticized for
`... [using] the interim estimate of the effect [...] twice ...' Bauer~\textit{et~al.},~p.~330~\cite{bauer2016} and \cite{bauer2006}.
Thirdly,
conditional power can be evaluated as Bayesian expected power conditional on the observed interim data,
i.e., by averaging conditional power as function of the unknown
parameters with respect to the posterior density
after conducting the interim analysis.
Within the hybrid Bayesian framework this is usually referred to as
`predictive power'~\cite{spiegelhalter1994,bauer2016}.

The purpose of this paper is to argue that pre-planned sample size adaptations
using methodology intended for unplanned interim analyses to react to within-trial interim data are inefficient and unnecessary if the original design was planned optimally.
We then discuss cases where an unplanned design adaptation might still be warranted and propose two criteria that allow such an adaptation that is
consistent with the original design.
We first review assumed \mbox{conditional-,} observed \mbox{conditional-,} and predictive power
from an estimation perspective.
For sake of simplicity,
we use a simple single-arm one-stage design to illustrate core
characteristics.
We then discuss the risks of constructing an
`adaptive' two-stage design by na\"{\i}vely applying
a conditional-power-based sample size recalculation rule and the
conditional error principle.
The drawbacks of this na\"{\i}ve approach directly lead to the
concept of optimal-two stage designs and
we build on ideas presented in \cite{brannath2004,pilz2019}, and
\cite{adoptrjss2020} to derive optimal designs for the situation studied in
this manuscript.
Without loss of generality, we focus on minimal expected sample size as optimality criterion.
Optimal two-stage designs, by definition, cannot be made more
efficient by sample size recalculation.
The optimality of such designs does, however, depend on the initial trial-external evidence that feeds into the planning assumptions.
This trial-external evidence might change during an ongoing study
and may thus mandate a sample size recalculation.
In Section~\ref{sec:unplanned-adaptation},
we discuss when an unplanned sample size recalculation is
reasonable,
how such a recalculation interacts with the concept of
optimal two-stage designs,
and the practical implementation.

In the following,
we assume that the interest lies in testing a new treatment for efficacy.
For the sake of simplicity,
we consider the case of a single-arm trial.
All considerations can easily be extended to the more
practically relevant two-arm case.
We further assume that the individually observed outcomes of the study participants $X_i, i = 1,\ldots,n$ are $iid$ and that their distribution has finite first moment $\theta$ and unit variance.
Again, all considerations can be extended to the case of
generic known variances and,
at least approximately, to the case of unknown variance.
A suitable test statistic for the null hypothesis of interest~${\mathcal{H}_0:\theta\leq0}$ is
\begin{align}
    Z_n = \frac{1}{\sqrt{n}} \sum_{i=1}^n X_i \ .
\end{align}
Invoking the central limit theorem,
$Z_n\stackrel{\cdot}{\sim}\mathcal{N}(\sqrt{n}\,\theta, 1)$
and,
on the boundary of the null hypothesis, $Z_n\stackrel{\cdot}{\sim}\mathcal{N}(0, 1)$.

Further assuming a maximal permissible type-I error rate of $\alpha=0.025$ (which we use for the remainder of the paper), the critical value for a single-stage fixed-sample-size design is the
$1-\alpha$ quantile of the standard normal distribution, i.e., approximately $c=1.96$.
The test then rejects $\mathcal{H}_0$ if and only if $Z_n > c$ after the outcomes of $n$ subjects have been observed.
The required size of the trial $n$, given a maximal allowable type-I error
rate, is usually determined by some form of restriction on
the (minimal) statistical power of the test.
Approaches to defining such a power constraint under
different assumptions were reviewed in~\cite{kunzmann2020}.

\section{Monitoring power} %%%%%%%%%%%%%%%%%%%%%%%
\label{sec:monitoring-power}

No matter which rationale was used to derive the overall sample size $n$,
after observing $m$, $0<m<n$, outcomes,
an independent data monitoring committee might be interested to learn about the prospects of
eventually rejecting the null hypothesis.
Both the rejection probability conditional on the presence of an effect (conditional power) or the joint probability of rejection and presence of an effect (conditional probability of success) could be of interest.
Here, we focus on the former due to the central role of
power (and conditional power) in the planning of clinical trials.
This effectively constitutes an estimation problem where the estimand is
the \emph{conditional} probability of rejecting the null hypothesis given
$Z_m=z_m$ and $\theta > 0$.
Here the conditioning on a positive effect is crucial to justify the notion
of `power' in contrast to the unconditional `probability of success'~\cite{kunzmann2020}.

Since $Z_n$ and $Z_m$ are (asymptotically) jointly normal, the conditional distribution of $Z_n$ given $Z_m$ and $\theta$ is again normal and
given by
\begin{align}
    \mathcal{L}_\theta\big[\, Z_n \mid Z_m = z_m \,\big] \stackrel{\cdot}{=} \mathcal{N}\Big(\sqrt{n}\,\theta + \sqrt{\tau}\big(z_{m} - \sqrt{m}\,\theta\big),\ 1 - \tau\Big) \label{eq:joint-distribution}
\end{align}
where $\tau:=m/n$ is the `information fraction' at the interim analysis.

The probability of rejecting the null hypothesis at the end of the trial
as a function of $\theta$ is referred to as the `conditional power' in the literature.
It is defined as
\begin{align}
    \CP(z_m, c, \theta) :&= \Pr_\theta[\,Z_n > c \cond Z_m = z_m\,] \\
        &= 1 - \Phi\left(
            \frac{c - \sqrt{n}\,\theta - \sqrt{\tau} \, z_m + \sqrt{\tau}\sqrt{m}\,\theta
            }{\sqrt{1 - \tau}}
        \right) \\
        &= 1 - \Phi\left(
            \frac{c/\sqrt{m} + \sqrt{\tau}\,\big(\theta - \widehat{\theta}_m\big) - \theta/\sqrt{\tau}
            }{\sqrt{(n - m)/(n\,m)}}
        \right)      \ .
\end{align}
Ignoring the implicit dependency on the time point of the interim analysis, $m$,
conditional power is thus a function of the observed effect $\widehat{\theta}_m:=z_m/\sqrt{m}$
(or equivalently the observed test statistic $z_m$), the critical value for the test decision~$c$,
and the standardized effect size~$\theta$.
Since $\theta$ is unknown,
so is $\CP(z_m, c, \theta)$ and it thus cannot be
evaluated directly upon observing $Z_m=z_m$.
Yet, as a function of the unknown quantity $\theta$,
$\CP(z_m, c, \theta)$ can be estimated from observed data.
The estimation perspective on `evaluating' conditional power is
less commonly taken in the literature but provides a consistent framework to
compare the characteristics of different methods~\cite{bauer2006}.

Firstly, conditional power can be estimated based on a fixed point alternative~$\theta_1$.
In a slight abuse of terminology,
this quantity is often also referred to as `conditional power'.
To clearly distinguish it from $\CP(z_m, c, \theta)$ we denote
it `assumed conditional power'
\begin{align}
    \ACP(z_m,c) := \CP\big(\, z_m, c, \theta_1 \,\big) \ .
\end{align}

Secondly, the observed effect $\widehat{\theta}_m$ can be used
as a plug-in estimator for the unknown effect size~\cite{proschan1995}.
This quantity is sometimes referred to as `observed conditional power'
in the literature and is defined as
\begin{align}
    \OCP(z_m, c) := \CP\big(\, z_m, c, \widehat{\theta}_m \,\big) \ .
\end{align}

Thirdly, a Bayesian approach can be taken if one
is willing to quantify the uncertainty about the unknown parameter $\theta$ by modeling it as the
realization of a random variable $\Theta\sim\varphi(\cdot)$ where $\varphi(\theta)$ is the
prior probability density function evaluated at the parameter value $\theta$.
Our definition of this so-called `predictive power'
\begin{align}
    \PP_\varphi(z_m,c) :&= \Pr_\varphi[\,Z_n > c \cond \Theta >0, Z_m = z_m\,] \\
    &= \int_{0}^{\infty}\CP\big(z_m, c, \theta\big)\,\varphi\big(\theta\cond Z_m=z_m, \Theta\geq0\big)\operatorname{d}\theta
\end{align}
differs slightly from the one proposed by~\cite{spiegelhalter1994}
in that we condition on a positive effect size $\Theta>0$.
This is more consistent with the notion of frequentist power as discussed in~\cite{kunzmann2020}, although the difference is
only of practical relevance when a substantial
fraction of the \textit{a~priori} probability mass is concentrated on the null hypothesis.

\begin{figure}
    \centering
    \includegraphics[width=\textwidth]{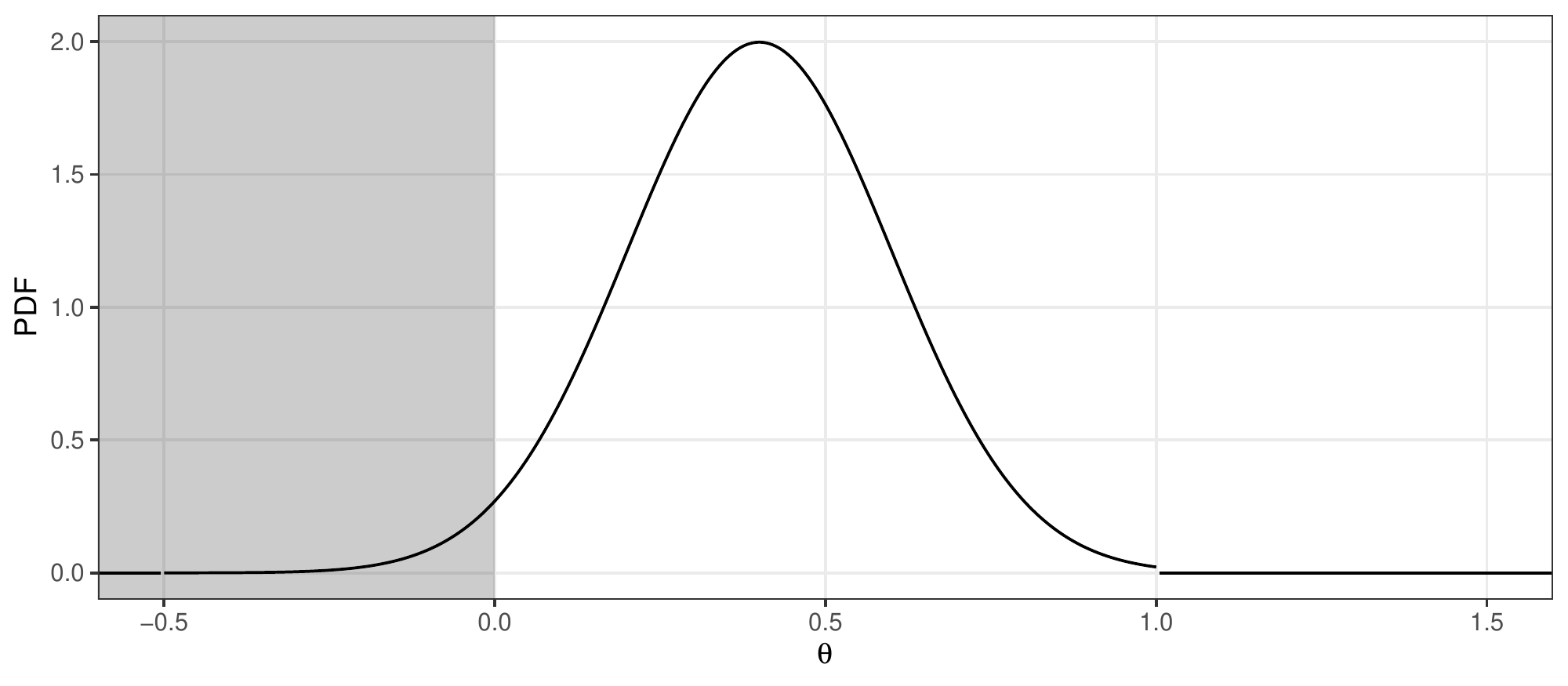}
    \caption{%
        Assumed prior density function.
        The gray area indicates the null hypothesis, $\mathcal{H}_0$, of no effect.
    }
    \label{fig:prior-pdf}
\end{figure}

We now compare $\ACP, \OCP$, and $\PP$ by means of a concrete example.
To this end we assume that the available prior information can be summarized in a truncated normal prior with density
\begin{align}
    \varphi(\theta) := \boldsymbol{1}_{[-0.5, 1]}(\theta)\,\frac{\displaystyle \phi\left(\frac{\theta - 0.4}{0.2}\right)}{\displaystyle \Phi\left(\frac{1 - 0.4}{0.2}\right) - \Phi\left(\frac{-0.5 - 0.4}{0.2}\right)} \ ,
\end{align}
(see Figure~\ref{fig:prior-pdf})
and that all positive effect sizes are clinically relevant.
A truncated normal prior is convenient since it allows the analytic
computation of the posterior distribution under a normal likelihood.
It is also the maximum entropy distribution on a compact interval given mean and standard deviation and thus `least informative'
given a lower and upper boundaries on plausible effects as well as
the location and vagueness (variance) of the prior.
Following~\cite{kunzmann2020} the required
sample size $n$ is then determined by requiring
a minimal expected power of $1-\beta=80\%$,
where expected power is defined as
\begin{align}
    \EP_\varphi(n, c) :=
    \Pr\big[\,Z_n>c\,|\,\Theta\geq0\,\big] = \int \Pr_\theta\big[\,Z_n>c\,\big]\, \varphi(\theta\,|\,\Theta\geq0) \,\operatorname{d}\theta \ .
\end{align}
Under these assumptions, the required sample size is
$n = 79$.
No point alternative is necessary to derive this sample size.
We thus use the prior mean rounded to the first decimal point to
evaluate $\ACP$ and define $\theta_1:=0.4$.

\begin{figure}
    \centering
    \begin{subfigure}[b]{\textwidth}
        \centering
        \includegraphics[width=.85\textwidth]{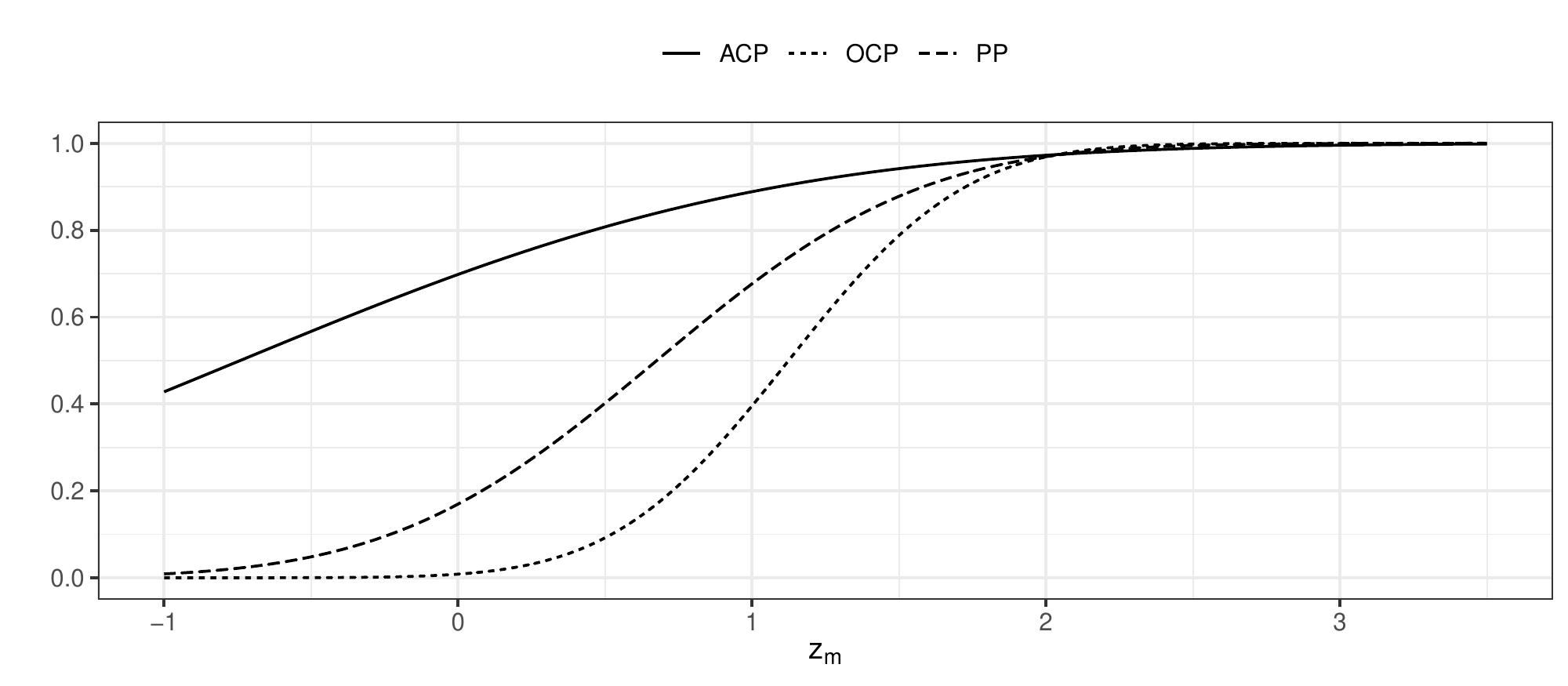}
        \caption{%
            Estimates as function of the interim data.
        }
        \label{fig:acp-ocp-pp-comparison-absolute}
    \end{subfigure}
    \begin{subfigure}[b]{\textwidth}
        \centering
        \includegraphics[width=.85\textwidth]{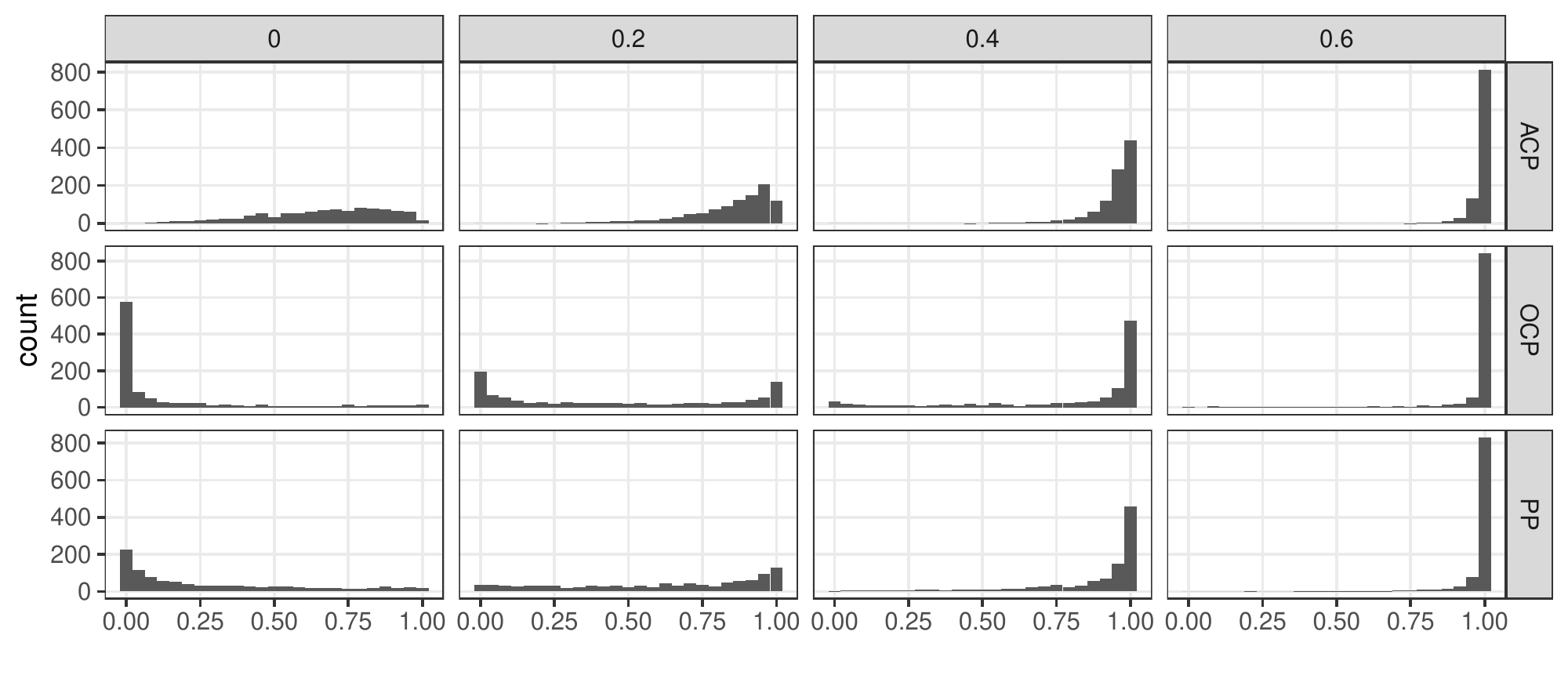}
        \caption{%
            Histograms of the sampling distributions.
        }
        \label{fig:sampling-distributions}
    \end{subfigure}
    \begin{subfigure}[b]{\textwidth}
        \centering
        \includegraphics[width=.85\textwidth]{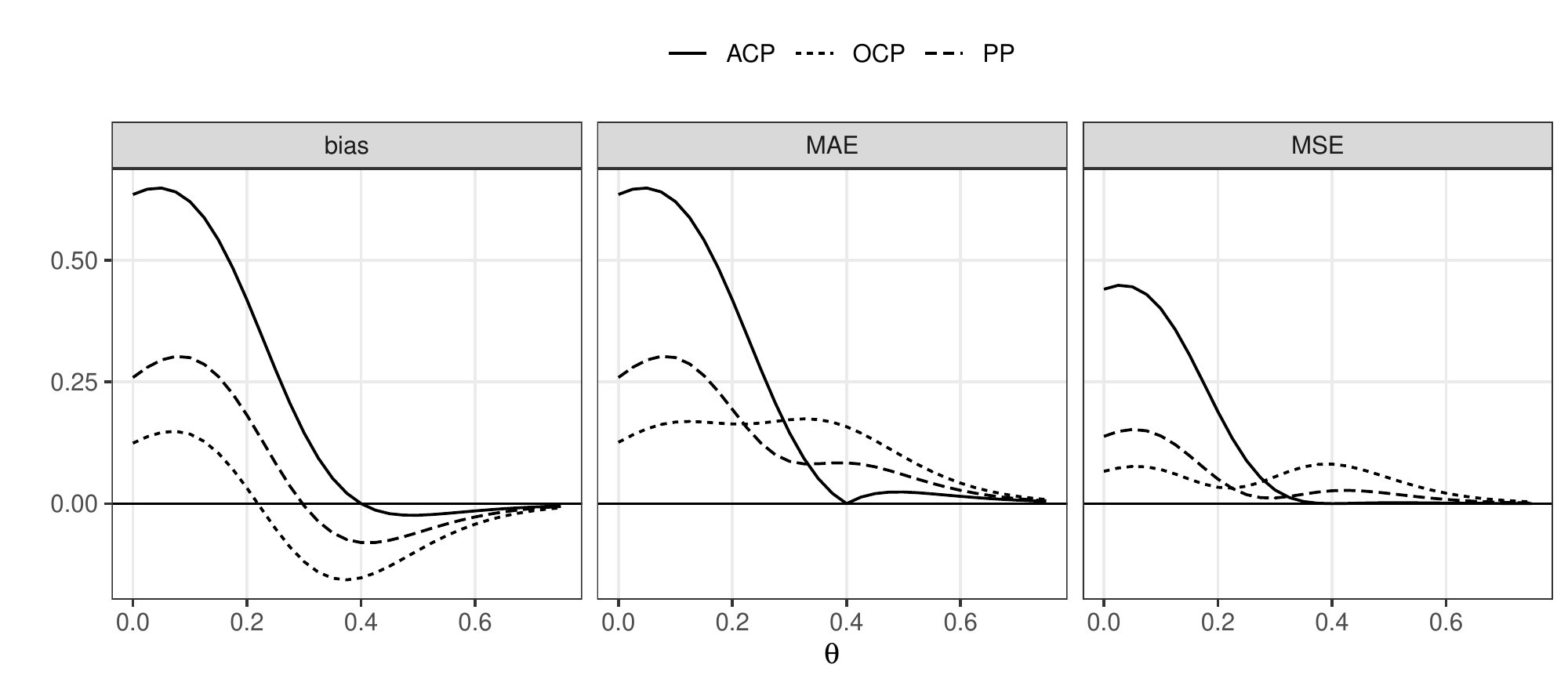}
        \caption{%
            Bias, mean absolute error (MAE) and mean squared error (MSE).
        }
        \label{fig:acp-ocp-pp-as-estimators}
    \end{subfigure}
    \caption{%
        Properties of $\ACP,\OCP$, and $\PP$ as estimators of the unknown conditional power at $m=26$ an overall sample size $n=79$.
    }
\end{figure}

The three estimators are depicted in Figure~\ref{fig:acp-ocp-pp-comparison-absolute} as functions
of the observed interim outcome, where the interim time-point $m = 26$.
The most sensitive measure is $\OCP$ while $\ACP$ is the least sensitive
(see also spread of the sampling distributions in Figure~\ref{fig:sampling-distributions}).
This is due to $\ACP$ assuming a fixed effect size and thus
only being affected by direct changes in $z_m$.
Both $\OCP$ and $\PP$, however, are also indirectly affected by updating
the belief about the effect size with the interim results.
The plug-in estimate $\OCP$ assumes that the observed effect is the
true effect and does not quantify the uncertainty around this value in any way.
$\PP$, on the other hand, invokes Bayes' theorem and then integrates
$\CP(z_m,c,\theta)$ with respect to the obtained posterior distribution.
The degree of adaptation of $\PP$ thus depends
on the vagueness of the chosen prior.
In fact, as the prior approaches a point mass at $\theta_1$, $\PP$ converges
to $\ACP$.
$\ACP$ can thus be seen as a mere special case of $\PP$ with a point prior
on $\theta_1>0$.

A fundamental difference between $\ACP$ and $\PP$ on the one hand and $\OCP$ on the other hand is that
$\OCP$ is the only estimator that does not condition on~$\Theta$.
In contrast, $\ACP$ implicitly conditions on $\Theta=\theta_1>0$ and
$\PP$~on~${\Theta > 0}$.
The practical implication of this can be seen
by looking at the sampling distributions of the
three estimators under different assumptions
about the value of $\theta$ (see~Figure~\ref{fig:sampling-distributions}).
Since $\OCP$ does not condition on $\Theta>0$, the sampling distribution is much more left-skewed for small effects than for the other two estimators.
Also, the negligence of the sampling variation of $\widehat{\theta}_m$, which $\OCP$ plugs in the expression for conditional power, leads to a larger variance for intermediate values of $\theta$ and the characteristic
U-shape \cite{bauer2006}.
This high variance of $\OCP$ directly translates to a relatively large mean absolute error (MAE) and mean squared error (MSE) when estimating conditional power,
see Figure~\ref{fig:acp-ocp-pp-as-estimators}.
$\PP$ is the posterior expectation of conditional power with respect to the chosen prior and thus  minimizes the quadratic Bayes risk, i.e.~the MSE.
This leads to relative good precision for parameter values with high \textit{a~priori} likelihood
(around $\theta=0.4$).
If one wanted to minimize the MAE directly,
the posterior median would be optimal.
The principle, however, remains the same and
the posterior mean is more consistent with the derivation of the initial sample size using expected power.
$\ACP$ is clearly the best in terms of bias and MAE/MSE for values close to or
above $\theta_1$, but its performance as estimator of the conditional power
quickly deteriorates for small effect sizes.

$\ACP$ is merely an extreme special case of $\PP$.
$\OCP$ is both hard to justify theoretically and has inferior precision
for parameter values with a high \textit{a~priori} likelihood.
$\PP$ is thus the natural choice for monitoring power during the course of a trial.
Depending on the prior chose, it's properties can be either more similar to $\ACP$ (low prior variance) or $\OCP$ (high prior variance).

\section{Na\"{\i}ve unplanned sample size adaptations}
\label{sec:naive-adaptations}

Monitoring the predictive power of an ongoing trial naturally raises the question of whether this information can be leveraged to improve the operating characteristics of a trial.
Typically, sample size recalculation is considered in the context of group-sequential designs at the pre-specified interim analysis \cite{proschan1995,brannath2004,bauer2016}.
From a statistical perspective,
there is no reason why a sample size recalculation should not be
performed for a study that initially was planned using a single-stage design.
For the sake of simplicity the following considerations thus assume a simple
single-stage design with fixed $n$ and $c$ as starting point and are
restricted to a single interim analysis.
All results can be generalized to the case of more complex starting designs and multiple interim analyses.

A method for sample size recalculation that is often discussed in the literature is to derive a new sample size $n'$ and a new critical value $c'$ such that some estimate of conditional power exceeds a threshold $1-\beta_{cond.}$.
Different choices for the estimator of conditional power
and the choice of $\beta_{cond.}$ have been discussed in the literature \cite{proschan1995,brannath2004,bauer2016}.
Strict overall type-I error rate control can be maintained by invoking the conditional error principle \cite{muller2001,muller2004,brannath2012}, i.e., by limiting the maximal type-I error rate of the new design to the maximal conditional type-I error rate of the original design given the observed $z_m$.
Often, trial protocols leave the exact choice of $1-\beta_{cond.}$ open since it can be chosen \textit{ad~hoc}
without compromising strict type-I-error rate control.
The basic concept of readjusting the sample size to achieve the desired conditional power is, however, a common approach, see for instance,~\cite{bhatt2013} and~\cite{mehta2009}.

In accordance with the discussion in the previous section
and to be consistent with the derivation of the
initial sample size of $n=79$ in the example considered earlier,
we use predictive power as an estimator of conditional power.
Furthermore, we set $\beta_{cond.}=\beta=0.2$.
For given $Z_m = z_m$, the recalculation rule then corresponds to solving
the optimization problem
\begin{align}
    \argmin{n', c'} &&                   n' & \label{eq:naive-recalc}\\
    \st             && \CP(m, n', z_m, c', 0) &\ \leq \ \CP_n(m, n, z_m, c, 0) \label{eq:ce-principle-naive-recalc} \\
                    &&    \PP_\varphi(m, n', z_m, c') &\ \geq \ 1 - \beta_{cond.} \\
                    &&                   n' &\ \geq \ n_{min} > m \label{eq:nmin-naive-recalc} \\
                    &&                   n' &\ \leq \ n_{max} \label{eq:nmax-naive-recalc} \ .
\end{align}
Here, we indicate the dependency on the
final sample size and the interim time point explicitly by redefining
\begin{align}
    \CP(m, n, z_m, c, \theta) :&= \Pr_\theta[\,Z_n > c \cond Z_m = z_m\,] \\
    \PP_\varphi(m, n, z_m, c) :&= \Pr_\varphi[\,Z_n > c \cond \Theta >0, Z_m = z_m\,] \ .
\end{align}
Constraint~\eqref{eq:ce-principle-naive-recalc} implements the conditional
error principle by limiting the conditional (type-I) error rate under the new design to
the conditional (type-I) error rate under the original design.
The minimal sample size constraint~\eqref{eq:nmin-naive-recalc} allows  $n_{min}>m$ for cases where a minimal sample size upon rejection of the null hypothesis is deemed necessary.
This could e.g.~be of interest if a minimal precision of the final maximum likelihood estimator is sought.
Note that the trial cannot stop at $n'=m$ without immediately accepting the
null hypothesis since the conditional error under the new design in the case of early acceptance would be $1$ but is always smaller than $1$ for the original single-stage design.
Imposing a maximal sample size of $n_{max}$ via constraint~\eqref{eq:nmax-naive-recalc} is a practical necessity since the
recalculated sample size could otherwise tend to infinity as $z_m$ approaches negative infinity.
In cases where this constraint prevents a solution (low observed effect),
the trial is usually stopped early for futility declaring that the null hypothesis cannot be rejected.
Alternatively, one could continue with the maximal sample size accepting
the fact that the conditional power constraint is not met.
This would, however, raise the question as to why a hard constraint on conditional power was imposed in the first place.

\begin{figure}
    \centering
    \includegraphics[width=\textwidth]{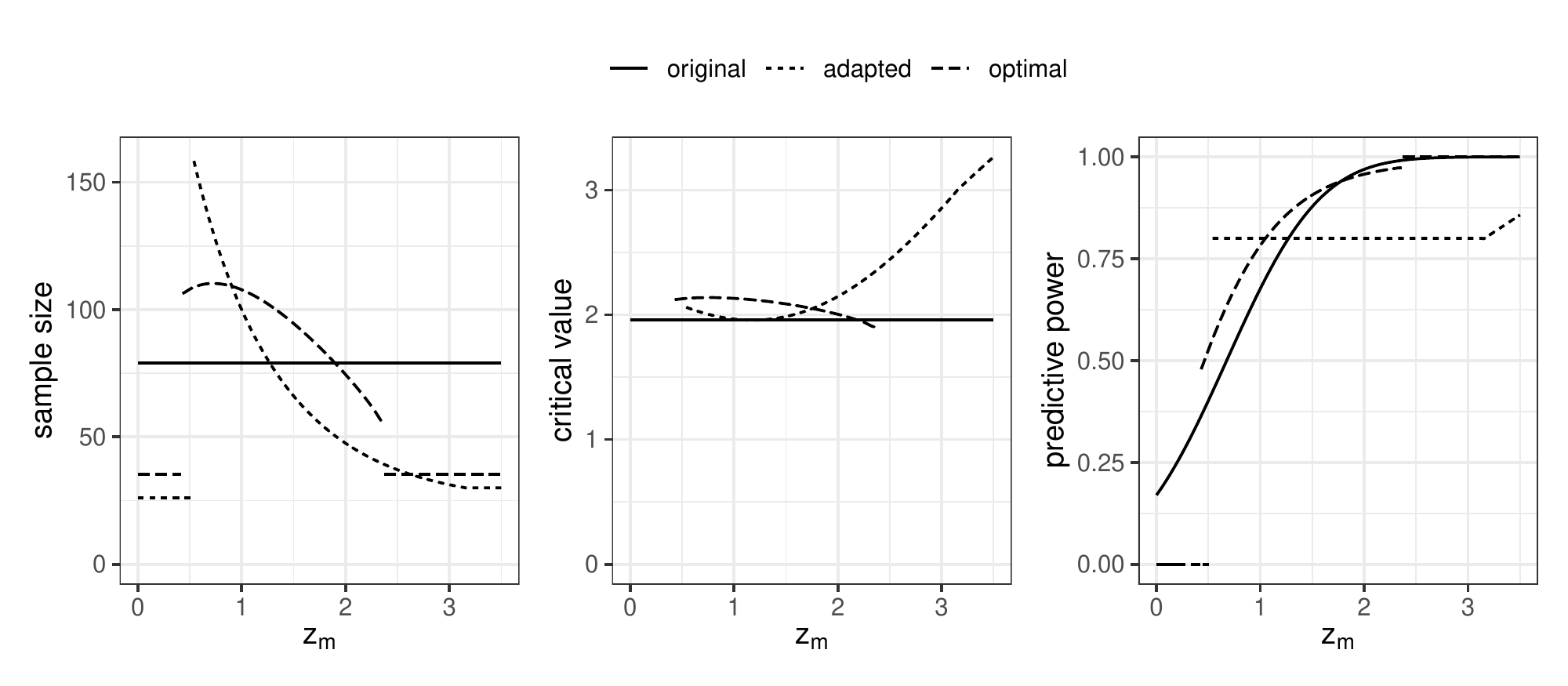}
    \caption{%
        Original single-stage design; na\"{\i}vely adapted design
        with sample size $n'(z_m)$ and critical value $c'(z_m)$ functions defined by the conditional optimization problem \eqref{eq:naive-recalc}--\eqref{eq:nmax-naive-recalc} and
        ${n=79}, {m=26}, {n_{max}=160}, {n_{min}=30}$ and $1-\beta_{cond.}=0.8$;
        the optimal design is the solution of \eqref{eq:optimal-design-start}--\eqref{eq:optimal-design-end}.
    }
    \label{fig:binding-recalculation}
\end{figure}

If the recalculation rule defined in \eqref{eq:naive-recalc}--\eqref{eq:nmax-naive-recalc} is made mandatory in the
study protocol,
it yields two functions $n(z_m)$ and $c(z_m)$ as the point-wise solution of the problem
which jointly define an `adaptive design' for any $z_m$.
Here, `mandatory' means that it is decided \textit{a~priori} to always (for all $z_m$)
recalculate the
final sample size and critical value in the
above specified way after a fixed number of outcomes $m$ has been observed.
The term `adaptive design' is slightly misleading in this context.
The design itself is pre-specified (and thus not adapted or changed)
but only $n$ and $c$ are adaptive since they vary as functions of $z_m$.
We would like to emphasize that such a detailed specification of the sample size recalculation rule is not common in practice since most protocols are focused on guaranteeing control of the maximal type-I-error rate control and maintaining maximal flexibility during the interim analysis.
Still, if a sample-size reassessment was prescribed in the study protocol, there must have been some rationale as to the goal of that reassessment and we merely investigate the consequences for a design's unconditional properties when applying such a rationale in a mandatory way.
The described approach should thus be seen as a hypothetical example highlighting the ineffectiveness that can arise from a na\"{\i}ve implementation of a recalculation based on a fixed threshold for
conditional power.
The corresponding sample size and critical value functions for $m=26$ are depicted in Figure~\ref{fig:binding-recalculation}.

The mandatory application of the adaptation rule
results in a two-stage design.
The final sample size is thus a random variable $n(Z_m)$.
This implies that the objective criterion of `minimal sample size' is no longer meaningful.
Instead, one could use a weighted sum, $\E[\,n(Z_m)\,] + \eta \sqrt{\V[\,n(Z_m)\,]}$, of the expected value and standard deviation of the sample size as the objective criterion.
The rationale for adding a penalty depending on the standard deviation of the sample size is that it might incur additional costs due to more
complicated logistics (e.g., on demand production of additional drug doses).
For the single-stage design this proposal always reduces to the fixed sample size $n$ since the standard deviation is 0.
Whether or not the two-stage design is considered better then depends on the choice of $\eta$,
i.e, the relative weight of the the standard deviation in the objective.
In the particular situation considered here, the expected sample size is $49.5$ and its standard deviation is $29.7$.
Since the original $n$ was $79$, the two-stage
design would be considered `better'
than the single-stage design for $\eta>0.99$.

This comparison, however, ignores the fact that
the original unconditional constraint of and expected power of more than $80\%$
is no longer fulfilled by the design with mandatory
sample size adaptation.
Both expected power ($72.1\%$ \emph{versus} $80.0\%$) and maximal type-I error rate ($2.1\%$ \emph{versus} $2.5\%$) are lower than for the original design.
The difference in operating characteristics is mainly due to the fact
that the new design implicitly introduces a binding early-stopping boundary for futility
when the recalculation problem does not yield a $n'\leq n_{max}$.
This effect is less pronounced when the same recalculation rule is applied
to an initial two-stage group-sequential design
which already defines binding early stopping boundaries
for futility and efficacy.

Similar to suggestions in the literature on binding futility stopping,
one could modify the initial nominal $\alpha$ and the conditional power threshold $\beta_{cond.}$ until the recalculated design again fulfills the required
operating characteristics \cite{brannath2004}.
This approach to define a fully pre-specified two-stage design is, however, unnecessarily complicated in that it
uses methods like the conditional error principle or a combination function, which are originally intended for \emph{unplanned}
design adaptations.

\section{Optimal pre-planned sample size adaptations}
\label{sec:optimal-adaptation}

Evidently, the na\"{\i}ve application of the predictive power
adaptation rule outlined earlier is needlessly conservative.
Fixing the operating characteristics by tuning the nominal $\alpha$ and
$\beta_{cond.}$ does not address the deeper issue that the sample size and
critical value functions are still derived based on a merely heuristic recalculation criterion and that
they still implicitly depend on an original design that is never actually realized via the
conditional error constraint~\eqref{eq:ce-principle-naive-recalc}.

Instead, the interim time-point $m$,
the sample size function $n(z_m)$,
the critical value function $c(z_m)$,
and the early futility- and efficacy boundaries
($n(z_m)=m$ and $ c(z_m)=\pm\infty$)
can be optimized directly.
For the sample size function $n(z_m)$ \cite{brannath2004} discussed this approach by fitting a polynomial of degree 4 but they did not optimize
any of the other parameters.
In a more general form, \cite{pilz2019} approached the problem using
variational methods.
An \texttt{R} implementation that uses cubic splines for both the $n$ and
$c$ functions is available via the package~\textit{adoptr}~\cite{adoptrjss2020}.

In the following we restrict the considerations to the case
where only expected sample size is of interest and the variation is
ignored, i.e.~$\eta=0$.
The corresponding optimal two-stage design is the solution of
\begin{align}
    \argmin{m, n(\cdot), c(\cdot)}
        &&                                   \E_\varphi[\,n(Z_m)\,] & \label{eq:optimal-design-start}\\
    \st &&                 \Pr_{0}[\,Z_{n(Z_m)} > c(Z_m)\,] & \leq \alpha \\
        &&     \EP_\varphi\big(n(\cdot), c(\cdot)\big) & \geq 1 - \beta \label{eq:optimal-design-end}
\end{align}
and can be derived numerically using the \texttt{R} package \texttt{adoptr}~\cite{adoptrjss2020}.
Here, $\E_\varphi[\,n(Z_m)\,]$ is the expected sample size under the prior $\varphi$ and expected power needs to be redefined to account for the
fact that $n$ and $c$ are now functions of the interim results
\begin{align}
    \EP_\varphi\big(n(\cdot), c(\cdot)\big) :=
    \Pr_\varphi\big[\,Z_{n(Z_m)}>c(Z_m)\,|\,\Theta\geq0\,\big] \ .
\end{align}
Figure~\ref{fig:binding-recalculation} shows the sample size-, critical value-, and the predictive power function in the example situation discussed earlier together with the single-stage design
and the na\"{\i}ve adaptation based on predictive power discussed in
Section~\ref{sec:naive-adaptations}.
The optimal two-stage design complies with both the predictive power and the maximal type-I error rate constraints and is thus comparable with the original single-stage design in terms of sample size.
The expected sample size is much lower ($56.4$ \emph{versus} $79$) at the cost of a non-zero standard deviation of the expected sample size ($28.5$ \emph{versus} $0$).
Since it is entirely pre-specified,
the optimal two-stage design does not need to fall back to the conditional error principle to control the maximal type-I error rate.
Instead, it achieves the desired design characteristics in an optimal way since they are incorporated as constraints to problem \eqref{eq:optimal-design-start}--\eqref{eq:optimal-design-end}.
The optimal design fully exhausts the allowable maximal type-I error rate
and complies with the expected power constraint.
This comes at the cost of a predictive power that can drop as low as 40\% close to the futility boundary
as shown in the right panel of Figure~\ref{fig:binding-recalculation}.
The sample size function of the optimal design is also
characteristically different from the na\"{\i}ve design.
The latter is convex on the continuation region whereas the optimal
shape has a mode close to the early-futility boundary.
A recalculation based on exceeding a fixed lower threshold for
some estimator of conditional power always results in a convex
sample size function and can thus never be fully optimal irrespective of how the nominal values of $\alpha$ and $\beta_{cond.}$
are chosen.

The fact that the optimal two-stage design exhibits a monotonously increasing
predictive power rather than a constant predictive power of, say, 80\%
indicates that any recalculation rule keeping predictive power
close to a fixed target value must be inefficient in terms of minimizing expected sample size.
The mandatory application of the heuristic recalculation rule introduced in Section~\ref{sec:naive-adaptations} would change the optimal design for almost every value of $z_m$ and thus increase expected sample size
of the resulting design.
It may thus appear as if any `recalculation' of an optimal design's sample size using methods
for unplanned interim analyses was not only
unnecessary but even counterproductive.
This is, however, only the case if the initial design was optimal and
the planning assumptions still hold.

\section{Consistent unplanned recalculation}
\label{sec:unplanned-adaptation}

The preceding section made the case that the direct optimization of all design parameters is superior to the mandatory application of methods for unplanned design adaptations during the planning phase of a trial.
By definition,
no optimal design ever needs to be changed as a reaction to trial internal data during the course of the study.
However, there are still situations in which an unplanned sample size recalculation is warranted.
The optimality of a design typically depends on the chosen prior density $\varphi$ and the objective function.
At any point in time,
the prior density encodes all trial-external evidence
about the effect size.
This might change over the course of a trial.
For instance, another study might publish new results that trigger a reassessment of the considerations leading to the choice of $\varphi$.

In the following, we discuss two approaches of reacting to such trial-external events in a consistent way
if the original design minimizes expected sample size.
Here, `consistent' means that the original optimal design is invariant under
the recalculation rule unless the planning assumptions (prior) are changed or
the time-point of the interim analysis is different from the optimal one.
The considerations are restricted to cases where the result of an adaptation is a final sample size and no further interim analyses are planned for the remainder of the trial.
This means that the unplanned adaptation replaces the pre-planned interim analysis although it might occur at a different point in time.
All methods can be generalized to an unplanned interim analysis in the
first stage with a subsequent second interim analysis.
This might be necessary when the unplanned interim analysis has to
be conducted shortly after the start of a trial.

To obtain a consistent sample size recalculation rule,
the conditional error principle can be applied not only to the
maximal type-I error rate but also to the (average) type-II error rate.
Let $m'$ be the sample size of the unplanned interim analysis and
$\psi$ be the probability density function of the revised prior.
We propose to recalculate the new sample size, $n'$ and
the new critical value $c'$ as the solution of the point-wise optimization problem
\begin{align}
    \argmin{n',c'} && n' &\\
     \st && \CP(m', n', z_{m'}, c', 0) &\leq \Pr_0[\,Z_{n(Z_m)} > c(Z_m) \cond Z_{m'} = z_{m'}\,] \label{eq:conditional-error-recalc}\\
     && \PP_\psi(m', n', z_{m'}, c') &\geq \nonumber \\
     && & \hspace*{-1cm}\Pr_\varphi[\,Z_{n(Z_m)} > c(Z_m) \cond \Theta >0, Z_{m'} = z_{m'}\,] \label{eq:pp-recalc} \ .
\end{align}
Here, constraint~\eqref{eq:pp-recalc} means that the modified design under the new prior has at least as much predictive power as the original design under the original prior.
Constraint~\eqref{eq:conditional-error-recalc} implements the conditional error principle and thus ensures strict type-I error rate control of the
procedure.
The crucial difference to the procedure discussed in Section~\ref{sec:unplanned-adaptation} is the fact that the lower boundary on predictive power varies with $z_m$.

In cases where the interim analysis is conducted at the original $m$,
the problem simplifies to
\begin{align}
    \argmin{n',c'} && n' & \label{eq:objective-original-interim}\\
    \st
    && \CP(m, n', z_{m}, c', 0) &\leq \CP(m, n(z_m), z_{m}, c(z_m), 0) \\
    && \PP_\psi(m, n', z_{m}, c') &\geq \PP_\varphi(m, , n(z_m), z_{m}, c(z_m))  \label{eq:pp-original-interim} \ .
\end{align}
Since $\CP$ is monotone in $n'$, the solution is already uniquely
defined by the constraints, if it exists.
It follows directly that, for $m'=m, \psi=\varphi$,
the original design ${n'=n(z_{m})}$, ${c'=c(z_m)}$ fulfills both constraints with equality.
The original optimal design is thus indeed invariant under the
proposed recalculation rule if neither the planning prior not the interim time-point are changed.

Problem \eqref{eq:conditional-error-recalc}--\eqref{eq:pp-recalc} is in no way the only conceivable
means of deriving a consistently recalculated sample size.
For the sake of simplicity, we consider the case of $m'=m$ (i.e.~problem \eqref{eq:objective-original-interim}--\eqref{eq:pp-original-interim}).
Depending on how much the prior changes,
the required sample size adjustment to meet the predictive power constraint can be quite drastic.
In practice, it might thus be preferable to relax the predictive power constraint in a principled way.
This can be done via the Lagrangian mechanism.
Let
\begin{align}
    f(n', c') :&= n' \\
    g(n', c') :&= \CP(m, n', z_{m}, c', 0) - \CP(m, n(z_m), z_{m}, c(z_m), 0) \\
    h_\psi(n', c') :&= \PP_\varphi(m, , n(z_m), z_{m}, c(z_m)) - \PP_\psi(m, n', z_{m}, c')
\end{align}
As discussed earlier, the original designs $n$ and $c$ trivially are the unique solution of
\begin{align}
    \argmin{n',c'} && f(n',c') & \\
    \st
    && g(n',c') &\leq 0 \\
    && h_\varphi(n',c') &\leq 0 \ .
\end{align}
This problem is equivalent to solving the unconstrained problem
\begin{align}
    \argmin{n',c'} f(n',c') - \lambda_g g(n',c') - \lambda_h h_\varphi(n',c')
\end{align}
for suitable choice of the Lagrange multipliers $\lambda_g$ and $\lambda_h$.
Since the original $n, c$ are optimal,
$\lambda_g$ and $\lambda_h$ can be obtained directly as solution of
\begin{align}
    0 &= \nabla f(n, c) - \lambda_g \nabla g(n, c) - \lambda_h \nabla h_\varphi(n, c) \\
   \Leftrightarrow (1, 0)^\top &= \big(\nabla g(n, c), \nabla h_\varphi(n, c)\big) \cdot (\lambda_g, \lambda_h)^\top \\
   \Leftrightarrow (\lambda_g, \lambda_h)^\top &= \big(\nabla g(n, c), \nabla h_\varphi(n, c)\big)^{-1} \cdot (1, 0)^\top\ .
\end{align}
Here, $\nabla$ denotes the gradient as column vector.
The Lagrange multiplier corresponding to the predictive power constraint, $\lambda_h$, encodes the trade-off between predictive power and sample size of the original,
optimal design at $z_m$.

A recalculated sample size under a different prior $\psi$ can then be obtained by
keeping $\lambda_h$ fixed and minimizing
\begin{align}
    \argmin{n',c'} && n'&-\lambda_h\, h_\psi(n',c') \\
     \st && g(n', c') &\leq 0 \ .
\end{align}
This effectively relaxes the predictive power constraint by allowing the same trade-off
between sample size and predictive power that was implied by the original optimal design.
The conditional type-I error rate constraint is not relaxed to comply with the
conditional error principle.
This approach is closely related to the formulation of the optimal-design problem given in~\cite{jennison2015} and the utility perspective on sample
size derivation presented in~\cite{kunzmann2020}.
In contrast to the situation analyzed by \cite{jennison2015}, however,
$\lambda_h$ does depend on $z_m$.
This is due to the fact that expected sample size is formed with respect to the unconditional
prior while expected power uses the prior conditional on $\Theta>0$.

\begin{figure}
    \centering
    \begin{subfigure}[b]{\textwidth}
        \centering
        \includegraphics[width=.8\textwidth]{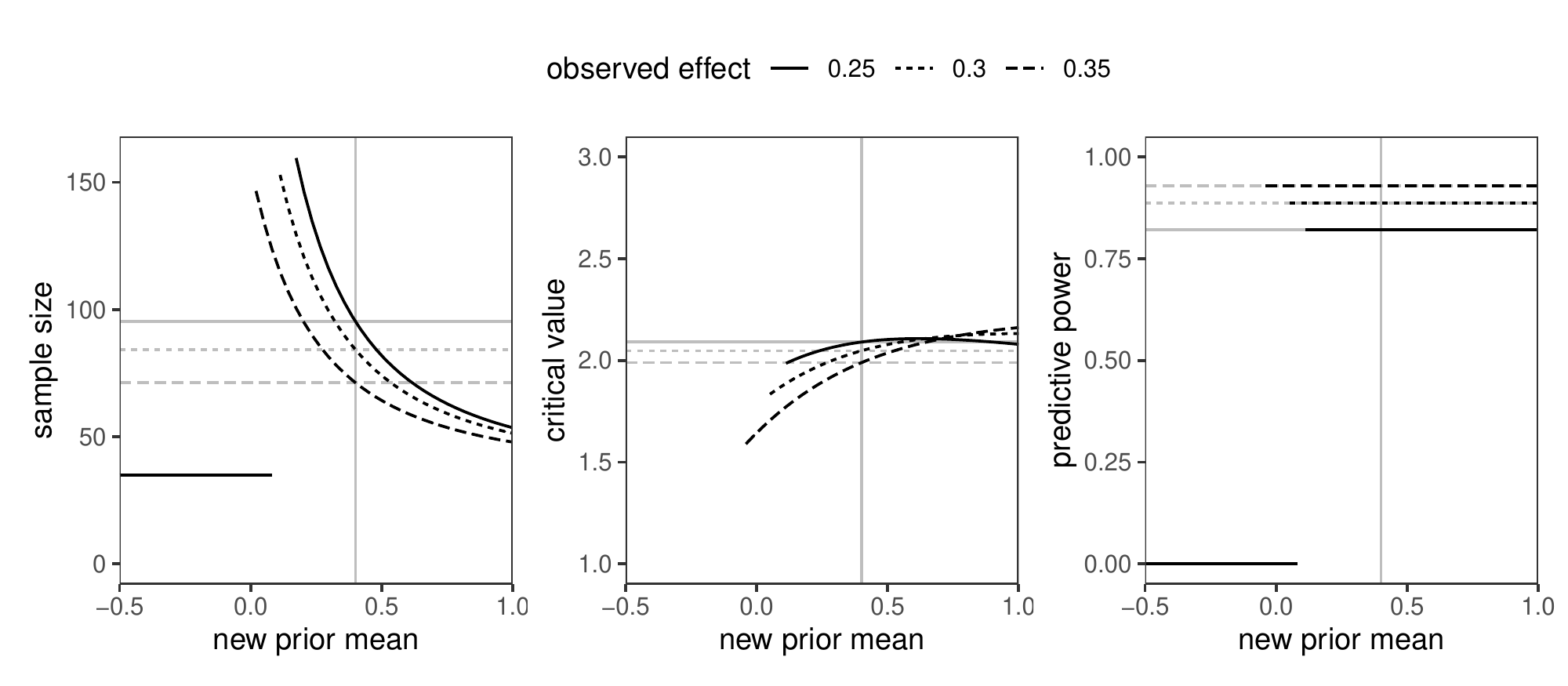}
        \caption{%
            Interim analysis at the originally planned interim sample size of $m=35$ for varying observed effect.
        }
        \label{fig:recalculation-old-m}
    \end{subfigure}
    \begin{subfigure}[b]{\textwidth}
        \centering
        \includegraphics[width=.8\textwidth]{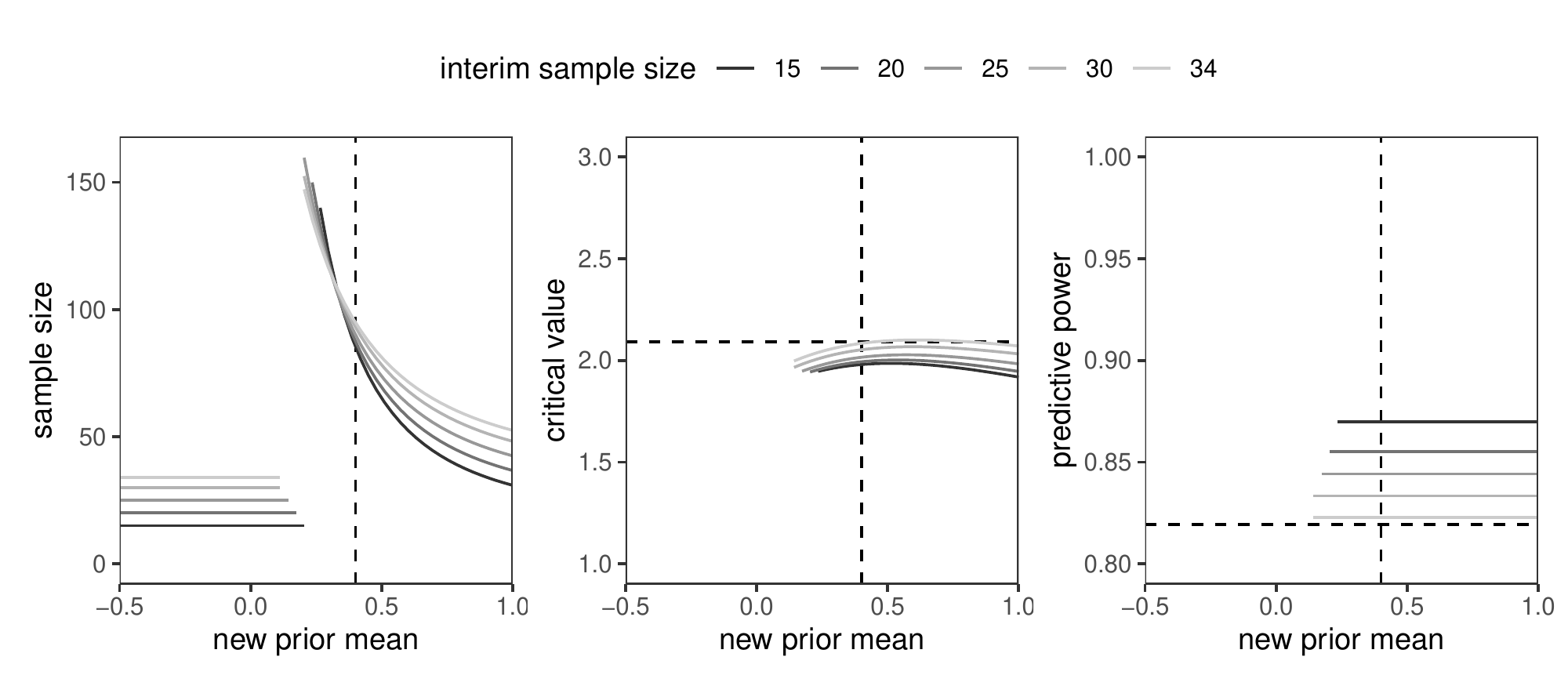}
        \caption{%
            Earlier-than-planned interim analysis at an observed effect of $0.3$ and varying interim time-point.
        }
        \label{fig:recalculation-earlier}
    \end{subfigure}
    \begin{subfigure}[b]{\textwidth}
        \centering
        \includegraphics[width=.8\textwidth]{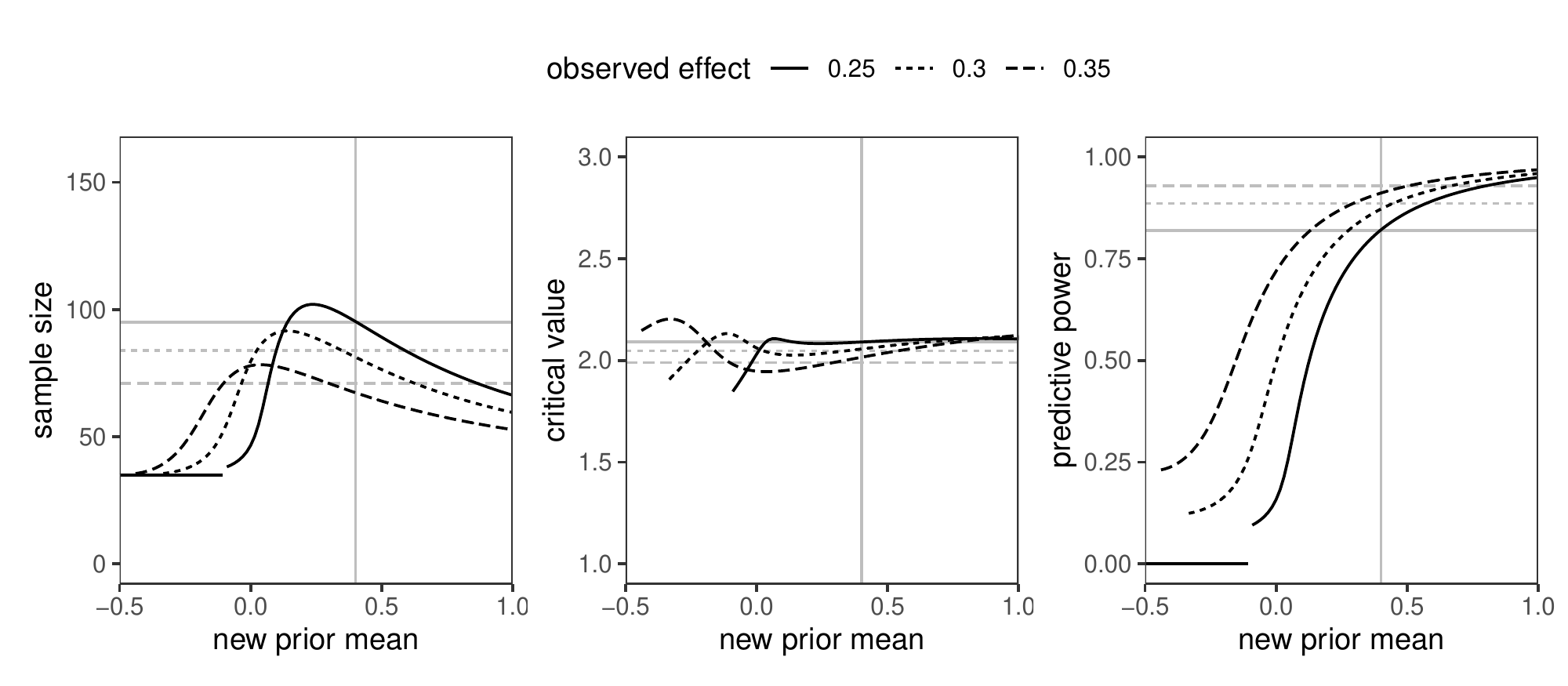}
        \caption{%
            Interim analysis at the originally planned interim sample size of $m=35$ for varying observed effect and the $\lambda$ approach.
        }
        \label{fig:recalculation-lambda}
    \end{subfigure}
    \caption{%
        Comparison of consistent unplanned recalculation schemes.
    }
\end{figure}

We illustrate both approaches by example of the optimal two-stage design that minimizes the objective
$\E_\varphi[\,n(Z_m)\,]$ subject to a maximal type-I error rate of $0.025$ and a minimal expected power of $0.8$, see~Figure~\ref{fig:binding-recalculation}.
This design has a pre-planned interim analysis after $m=35$ individuals.
Now assume that
the prior $\varphi$ needs to be revised due to new trial-external
information at the pre-planned interim analysis.
We consider members of the class of shifted truncated Normal distributions
\begin{align}
    \psi_{\mu}(\theta) := \frac{\displaystyle \boldsymbol{1}_{[-0.5, 1]}(\theta)\,\phi\left(\frac{\theta - \mu}{0.2}\right)}{\displaystyle \Phi\left(\frac{1 - \mu}{0.2}\right) - \Phi\left(\frac{-0.5 - \mu}{0.2}\right)}
\end{align}
as revised priors, i.e. the original prior~$\varphi=\psi_{0.4}$.
Figure~\ref{fig:recalculation-old-m} shows the behavior of the recalculated sample size and critical value in this scenario for observed effects of $z_m/\sqrt{m}=0.25, 0.3, 0.35$ and varying mean of the revised prior if the originally planned interim time-point is maintained.
Clearly, the original design is indeed invariant if the prior remains unchanged.
The sample size is increased if the new prior is more conservative
than $\varphi$ and \textit{vice~versa}.

Of course, it could also be necessary to conduct the interim analysis earlier or later than originally anticipated, at $m'\neq m$.
In this case, the design is no longer invariant since the earlier
or later time point already implies a deviation from the
original optimal design even if the prior remains the same.
Figure~\ref{fig:recalculation-earlier} shows the resulting adaptations
for earlier-than-planned time points of the interim analysis
while the observed effect is kept fixed at $0.3$.
The formula for the conditional error in this situation is given in the Appendix.

Figure~\ref{fig:recalculation-lambda} shows the recalculated sample size
for $m'=m=35$ and the $\lambda$-approach.
The original design is again (approximately) invariant under recalculation.
Compared to the fixed conditional type-II error approach, predictive power
under the $\lambda$-approach is more flexible and might drop below the predictive power under the original design, see Figure~\ref{fig:recalculation-lambda}, right panel.
This flexibility in terms of predictive power leads to a much smaller deviation from the original design's sample size.
The $\lambda$-approach transitions smoothly from
an increase of the original sample size due to a slightly more conservative prior to a decrease under very conservative priors.
This behavior leads to less extreme adjustments and is thus
more realistic in practice.
After all, a slight loss in (predictive) power as compared to the
original design might be tolerable if the sample size can be kept at
a moderate level in exchange.
Of course, it is still possible to stop the trial early for futility
at any point without compromising type one error rate control
if the predictive power is considered too low to warrant a continuation.

\section{Discussion} %%%%%%%%%%%%%%%%%%%%%%%%%%%%%%%%%%%%%%%%%%%%%%%%%%%%%%%

Monitoring the power of the remainder of an ongoing trial constitutes an estimation problem since the conditional power
is a function of the unknown treatment effect.
An often overlooked aspect of this estimation problem is that it is
conditional on a positive treatment effect since the notion of power is conditional on there being an non-null effect~\cite{kunzmann2020}.
In many practical situations where the prior mass is concentrated on
non-null effects the effect of conditioning is negligible but its importance increases with the vagueness of a the prior.
This lack of conditioning on a positive treatment effect is one of the reasons observed conditional power performs rather poorly as an estimator of the unknown conditional power.
Assumed conditional power for $\theta_1>0$ does reflect this conditionality.
It is, however,
merely a special case of predictive power with a
point prior on the effect size.
In situations where the effect size is fairy well known, a point prior might constitute an acceptable, simpler approximation.
In any other case, the Bayesian predictive power allows \textit{a~priori} information to be incorporated in a more fine-grained way and
guarantees optimal mean-squared-error performance.
If an unconditional measure for the probability to reject the null hypothesis is sought,
a conditional version of the probability of success (or assurance) can be derived in an analogue way~\cite{kunzmann2020}.

Great care should be taken when predictive power
or another estimator of conditional power
is used to modify an ongoing trial's sample size.
Altering a simple starting design with a seemingly intuitive recalculation rule
can actually lead to overall \emph{less effective} trials.
Techniques originally intended for \emph{unplanned} adaptations of trials
should not be used for a \emph{pre-planned} sample size recalculation.
Instead, an optimal two-stage design should directly be derived
for the objective criterion of interest.
By definition, no trial-internal event can then justify a sample size recalculation.
Only trial external events, such as a change in the objective criterion,
the emergence of new trial-external evidence,
or unforeseen deviations from the planned interim analysis time-point
may require a reassessment of the sample size of an optimal design.
A generic recalculation rule might lead to the paradoxical situation that even an optimal
starting design is always modified during the interim analysis
- even if the planning assumptions remain unaltered throughout the trial.
This is clearly ineffective and, consequently,
an optimal starting design should be invariant under a sensible adaptation rule
if the planning assumptions remain unchanged.
For instance, this minimal consistency property is not fulfilled when recalculating a design's
sample size based on a fixed threshold for its minimal conditional power
since designs optimizing expected sample size tend to have variable conditional power.

We propose two consistent approaches to adjusting an ongoing optimal design to
newly emerging trial-external data.
The first method applies the conditional error principle to the (average) type-II error rate in a
similar way to its use in controlling the maximal type-I error rate.
The method is easy to implement and consistent in the above defined way.
We would like to stress that this is by no means the only way of
conducting a consistent sample size recalculation.
The alternative $\lambda$-approach allows a trade-off between predictive power
and sample size leading to smaller sample size deviations from the original design.
This comes at the risk of potentially loosing predictive power.
Both methods differ by the quantity that is held fixed during the recalculation.
For the first method, the predictive power under the old design and the old prior
is maintained for the new design under the new prior whereas the $\lambda$-approach
only keeps the trade-off between predictive power and sample size fixed but allows the
actual predictive power to deviate form the original design.

These considerations show how absolutely crucial the initial planning stage of a trial is.
Adaptive methods should not be taken as an excuse to start with a
sub-optimal design and rely on a later sample size recalculation.
Ideally, all uncertainty is quantified during the planning phase to
the best possible extend and integrated in the initial design via a
planning prior.
This does not mean that a complex design is always the best choice.
If the variability of the final sample size and the operational
burden of conducting interim analyses is penalized strong enough in
the objective criterion, a simple one-stage design might very well
perform better than more complex alternatives.
Also,
optimal-two stage designs require optimizing over function-spaces to find the optimal sample size and critical value functions.
Obtaining a stable solution might thus be hard in practice and
group-sequential designs are a viable approximation since optimizing the stage-wise sample sizes and stopping boundaries only requires optimizing over a small set of real parameters.
It is well-known, that optimal group-sequential designs approximate the performance of optimal two-stage designs with variable sample sizes sufficiently well~\cite{wassmer2016}.
Still, the same principle considerations apply:
An optimal (group-sequential) design only needs to be revised if either the objective function or the underlying planning assumptions, i.e., the prior on the effect size, changes.

Adaptations of initially sub-optimal designs are another potential
application of the new recalculation scheme.
Clearly, the single-stage design with $n=79$ is not minimizing
\emph{expected} sample size.
To switch to a near-optimal adaptive design mid-trial,
one could derive the optimal two-stage design given the current best prior \textit{post-hoc} and
then use the $\lambda$-approach to recalculate the initial design while controlling the maximal
conditional type-I error rate in the usual way.
The resulting design will not be optimal from an unconditional perspective since the conditional error function of the initial, sub-optimal design has to be respected but the power-sample size trade-off during the interim analysis
will be similar to the optimal design due to the choice of $\lambda_h$.

A limitation of the proposed recalculation methods is the fact that they
rely on the conditional error principle for strict type-I-error rate control.
The conditional error principle can be difficult to extend to cases with nuisance
parameters \cite{gutjahr2011}.
In practice, a simple plug-in approach similar to the approach in blinded sample size reassessment might be viable but have not yet been investigated more thoroughly.
Alternatively, any other form of pre-specifying a combination function for the stage-wise p values might be used to control the type-I-error rate.
This combination function should then also be optimized over during the planning stage
to avoid inconsistencies.
In particular, the optimal combination function of a two stage-design minimizing expected sample size can be approximated with an inverse normal combination test~\cite{pilz2019}.

\subsection*{Data Availibility Statement}
Data sharing is not applicable to this article as no datasets were generated and
the research is entirely theoretical.
The source code is available at
\href{https://github.com/kkmann/unblinded-sample-size-adaptation}{github.com}
and
\href{https://doi.org/10.5281/zenodo.3925752}{zenodo.org}.

\bibliography{manuscript.bib}

\section*{Appendix}

The conditional error for an earlier-than-planned interim analysis ($m'<m$)
is
\begin{align}
    & \Pr_\theta\big[\,Z_{n(Z_m)} > c(Z_m) \cond Z_{m'} = z_{m'}\,\big] \\
    =& \int \CP_\theta\big(m,n(z_m),z_m, c(z_m), \theta\big)\  p_\theta(z_m \cond z_{m'}) \operatorname{d} z_m
\end{align}
where $p_\theta(z_m \cond z_{m'})$ denotes the probability density function of the outcome at $m$ given the partial observed outcome up to $m'$ (see equation~\eqref{eq:joint-distribution}).

\end{document}